\documentclass[aps,prd,twocolumn,floats,nofootinbib,longbibliography]{revtex4-1}
\usepackage[dvips]{graphicx}

\usepackage{amsmath}
\usepackage{braket}
\usepackage{amsfonts}
\usepackage[colorlinks=true]{hyperref}
\usepackage{hyperref}
\usepackage{xcolor}

\usepackage{booktabs}

\usepackage{aas_macros}

\usepackage{caption}
\usepackage{graphicx}
\usepackage{subcaption}

\usepackage{bm}

\def\aap{\textit{Astronomy and Astrophysics}}

\def\be{\begin{equation}}
\def\ee{\end{equation}}
\def\bea{\begin{eqnarray}}
\def\eea{\end{eqnarray}}

\def\Lm{\mathcal{L}_m}

\begin{document}
\title{Compact Objects in Entangled Relativity}

\author{Denis Arruga}
\affiliation{Laboratoire Astroparticule et Cosmologie (APC), Universit\'e de Paris, France}
\author{Olivier Rousselle}
\affiliation{Laboratoire Kastler Brossel (LKB), Sorbonne Universit\'e, ENS-PSL, Coll\`ege de France, CNRS, Paris, France}
\author{Olivier Minazzoli}
\affiliation{Artemis, Universit\'e C\^ote d'Azur, CNRS, Observatoire C\^ote d'Azur, BP4229, 06304, Nice Cedex 4, France}

\begin{abstract}
We describe the first numerical Tolman-Oppenheimer-Volkoff solutions of compact objects in entangled relativity, which is an alternative to the framework of general relativity that does not have any additional free parameter. Assuming a simple polytropic equation of state and the conservation of the rest-mass density, we notably show that, for any given density, compact objects are always heavier (up to $\sim 8\%$) in entangled relativity than in general relativity---for any given central density within the usual range of neutron stars' central densities, or for a given radius of the resulting compact object.

\end{abstract}
\maketitle

\section{Introduction}

Recently, a new theory of relativity has been proposed in which matter and geometry cannot be treated separately \cite{ludwig:2015pl,minazzoli:2018pr}. The reason being that in the Lagrangian that depicts the theory, matter and geometry are unequivocally related through a pure multiplicative coupling. In some sense, matter and geometry are entangled in this theory \cite{minazzoli:2018pr}, since it is not possible to turn one off in order to study the other: the very definition of the theory of relativity becomes intertwined to the definition of matter fields. Hence we shall name this theory \textit{entangled relativity}. 

One consequence of the pure multiplicative coupling is that one would expect that the theory predicts phenomena that are violently in contradiction with several aspects of general relativity, which are however thought to be in good adequation with observations and experiments. But it turns out that the theory's phenomenology often boils down to the one of general relativity due to an \textit{intrinsic decoupling} that arises at the level of the classical field equations---which had initially been found in the framework of scalar-tensor theories with a specific non-minimal scalar-matter coupling \cite{minazzoli:2013pr, minazzoli:2014pr, *minazzoli:2014pl, minazzoli:2016pr,minazzoli:2018pr}. In particular, deviations from general relativity in the solar system may only happen at the post-post-Newtonian order \cite{minazzoli:2013pr}---which is beyond current testing capabilities.

Nevertheless, the theory is not without its own open issues. Notably, it is still not clear whether or not the theory can explain the acceleration of the expansion of the universe \cite{minazzoli:2014pr, *minazzoli:2014pl,minazzoli:2018pr}---although a recent proposal goes in the right direction \cite{minazzoli:2020ds}---nor is it clear to which quantitative level the universality of free fall may be violated in this theory \cite{minazzoli:2016pr,minazzoli:2018pr}---despite the \textit{intrinsic decoupling} at the level of the classical field equations previously mentioned. Not to mention that the situation with respect to a quantum field theoretic treatment of such a pure multiplicative Lagrangian is entirely open and probably not without its own problems---whether or not the standard model paradigm for matter fields is still the most appropriate description of matter in this novel context.

One of the unusual aspects of the theory---with respect to general relativity and its usual modifications---is that there cannot exist such a thing as a vacuum solution of the field equations. For instance, a Minkowski space-time is not a solution of the field equations. What it implies at a practical level, more broadly, is that most of the simplifications that are usually used in order to study extreme spacetime solutions---such as black holes---cannot be used in entangled relativity. Nevertheless, it has recently been argued that vacuum solutions of general relativity can be accurate enough approximation of non-vacuum solutions in entangled relativity in a near vacuum situation outside the event horizon \cite{minazzoli:2020bh}---such that astrophysical black holes may be described by their usual mathematical idealisation like the Schwarzschild and Kerr solutions.

On the other hand, it seems that the standard tools and methods are still useful for less extreme objects---such as Neutron stars. In this manuscript, we therefore propose an exploratory study of static and non-rotating compact objects with the spherical symmetry through the use of the Tolman-Oppenheimer-Volkoff (TOV) framework, in order to get the first glimpse of what compact objects may look like in entangled relativity.

\section{Field equations}

The action of entangled relativity is defined by \cite{minazzoli:2018pr}
\be
S=-\frac{\xi}{2} \int \mathrm{d}_g^{4} x  \frac{\mathcal{L}_{m}^{2}}{R},
\ee
where
 $\mathrm{d}_g^{4} x \equiv \mathrm{d}^{4} x \sqrt{-g}$ is the space-time volume element,
the coupling constant $\xi$ has the dimension of $\kappa \equiv 8 \pi G / c^{4}$---where $G$ is the Newtonian constant and $c$ the speed of light---but not its value. In fact, $\xi$ does not appear in the classical field equations and therefore is purely related to the quantum field sector of the theory. It is yet to be discovered what value of $\xi$ would likely be consistent with our universe, if any. It is important to note that apart from $\xi$, the theory does not have any coupling constant related to the link between matter and geometry. Hence, at the classical level, entangled relativity has one parameter less than general relativity in order to describe the link between matter and geometry, in the sense that no parameter replaces the parameter $\kappa$ of general relativity at the classical level in entangled relativity \cite{minazzoli:2018pr}: the effective coupling that appears at the level of the field equation is dynamical.

For spacetimes that are such that $R \neq 0$, there is a one to  one correspondence at the classical level between the action of entangled relativity and a dilaton theory with the following action \cite{ludwig:2015pl,minazzoli:2018pr}

\be
\label{eq:sfaction}
S=\frac{1}{c} \frac{\xi}{\tilde \kappa} \int d_g^{4} x\left[\frac{\phi R}{2 \tilde \kappa}+\sqrt{\phi} \mathcal{L}_{m}\right],
\ee
where $\tilde \kappa$ is an effective coupling constant between matter and geometry, with the dimension of $\kappa$. The corresponding field equations read

\bea
&&G_{\alpha \beta}=\tilde \kappa \frac{T_{\alpha \beta}}{\sqrt{\phi}}+\frac{1}{\phi}\left[\nabla_{\alpha} \nabla_{\beta}-g_{\alpha \beta} \square\right] \phi,\label{eq:metr}\\
&&\frac{3}{\phi} \square \phi=\frac{\tilde \kappa}{\sqrt{\phi}}\left(T-\mathcal{L}_{m}\right), \label{eq:sceq}
\eea
and the conservation equation reads
\be
\label{eq:noncons}
\nabla_{\sigma}\left(\sqrt{\phi} T^{\alpha \sigma}\right)=\mathcal{L}_{m} \nabla^{\alpha} \sqrt{\phi},
\ee
with
\be
T_{\mu \nu} \equiv-\frac{2}{\sqrt{-g}} \frac{\delta\left(\sqrt{-g} \mathcal{L}_{m}\right)}{\delta g^{\mu \nu}}.
\ee

The \textit{intrinsic decoupling} discussed in the introduction comes from the fact that the right hand side of Eq. (\ref{eq:sceq}) is minute for matter fields that are such that $\mathcal{L}_m \sim T$, hence leaving the scalar field mostly unsourced, and therefore constant \cite{minazzoli:2013pr, minazzoli:2014pr, *minazzoli:2014pl, minazzoli:2016pr,minazzoli:2018pr}.

\section{TOV equations}

In order to derive the TOV equations, we assume that the system is static and has a spherical symmetry, such that we assume a metric with the following form
\be
\mathrm{d} s^{2}=-a(r) \mathrm{d} t^{2}+b(r) \mathrm{d} r^{2}+r^{2} \mathrm{d} \Omega^{2},
\ee
in unit $c = 1$, where $d\Omega^2$ stands for the $S^2$-sphere metric.
We further assume a perfect fluid matter content, whose energy-momentum tensor in a general frame reads
\be
\label{eq:tab}
T_{\beta}^{\alpha}=(\rho+P) u^{\alpha} u_{\beta}+P \delta_{\beta}^{\alpha},
\ee
where $\rho$ and $P$ are the total energy density and the pressure of the fluid respectively, related to the equation of state defined in \eqref{eq:EqS}.

\subsection{Metric's components and a definition of the mass}
First of all, we note as
\be
    D^\alpha_\beta=\frac{1}{\phi}(\nabla^\alpha\nabla_\beta-\delta^\alpha_\beta\square)\phi.
\ee
the dilaton energy-momentum tensor part in equation (\ref{eq:metr}).
The time-time component of the Einstein equations, leads to the following first order differential equation for $g_{rr}\equiv b$ metric component
\be
\dfrac{\dot{b}}{b}=\dfrac{1-b}{r}+rb\left(\dfrac{\Tilde{\kappa} \rho}{\sqrt{\phi}}-D^0_0\right), \label{eq:beq}
\ee
where
\be
D^0_0=\dfrac{1}{2b}\dfrac{\dot{\phi}}{\phi}\dfrac{\dot{a}}{a}+\dfrac{\Tilde{\kappa}}{3\sqrt{\phi}}(\Lm-T).
\ee
One can formally integrate equation (\ref{eq:beq}) for $b$, which leads to  
\be\label{eq:b}
b=\frac{1}{1-m \tilde{\kappa} / 4 \pi r},
\ee
where
\be
m(r) \equiv \frac{4\pi}{\tilde{\kappa}} \int_{0}^{r}\mathrm{d} r\left(\frac{\rho}{\sqrt{\phi}}
\tilde{\kappa}
-   \frac{1}{2 b} \frac{\dot{\phi}}{\phi} \frac{\dot{a}}{a}-\frac{\tilde{\kappa}}{3 \sqrt{\phi}}\left(\mathcal{L}_{m}-T\right)\right)r^{2}. \label{eq:m}
\ee
From \eqref{eq:b}, the ADM mass reads
\be
\mathfrak{M}_{ADM}=\lim\limits_{r \rightarrow \infty}m(r), \label{eq:mADM}
\ee
and does not necessarily coincide with the star mass $m_*$ defined by
\be
m_*=m(R_*),\hspace{0.2cm}\text{with }p(R_*)=0. \label{eq:massatrad}
\ee
Indeed, outside the star, matter energy density and pressure are negligible, so that they are set to zero and the relation between the ADM mass and the star mass reads
\be
\mathfrak{M}_{ADM}=m_*-\frac{2\pi}{\tilde{\kappa}}\lim\limits_{r \rightarrow \infty}\int_{R_*}^{r}\mathrm{d} r
\frac{ r^2}{ b} \frac{\dot{\phi}}{\phi} \frac{\dot{a}}{a}
\ee
On the other side, the r-r component of the Einstein equation yield a first order differential equation for $g_{tt}\equiv a$
\be
\frac{\dot{a}}{a}=\frac{b}{r}\left(\frac{P r^{2} \tilde{\kappa}}{\sqrt{\phi}}+1-\frac{1}{b}-\frac{2 r}{b} \frac{\dot{\phi}}{\phi}\right)\left(1+\frac{r}{2} \frac{\dot{\phi}}{\phi}\right)^{-1}.
\ee

\subsection{Equation for pressure}

Following Wald \cite{wald:1984bk}, we compute the pressure using the conservation equation from the diffeomorphism invariance
\be
\nabla_{\alpha} T_{\beta}^{\alpha}=\left(\delta_{\beta}^{\alpha} \mathcal{L}_{m}-T_{\beta}^{\alpha}\right) \partial_{\alpha} \log \sqrt{\phi} \equiv F_{\beta}
\ee
where the covariant derivative of a perfect fluid stress-energy tensor reads
\bea
\nabla_{\alpha} T_{\beta}^{\alpha}&=&\left(\delta_{\beta}^{\alpha}+u^{\alpha} u_{\beta}\right) \partial_{\alpha} P+(P+\rho) u^{\alpha} \nabla_{\alpha} u_{\beta} \nonumber\\
&&+u_{\beta}\left[u^{\alpha} \partial_{\alpha} \rho+(P+\rho) \nabla_{\alpha} u^{\alpha}\right].
\eea
Projecting the last equation on $\partial_r$ and using $(\partial_r)^\mu u_\mu = 0$ brings up a derivative of the pressure with respect to the radius and we end up with
\be
\dot{P}=-\frac{\dot{a}}{2 a}(P+\rho)+\frac{\dot{\phi}}{2 \phi}\left(\mathcal{L}_{m}-P\right).
\ee

\subsection{Equation for the scalar-field}

The scalar-field equation reduces to
\be
\ddot{\phi}=-\frac{\dot{\phi}}{2}\left[\frac{\dot{a}}{a}-\frac{\dot{b}}{b}+\frac{4}{r}\right]+\frac{\tilde{\kappa} \sqrt{\phi}}{3} b\left(T-\mathcal{L}_{m}\right).
\ee

At this point, the system can be fully inverted from $(a,b,\phi)$ to $(m, P, \phi)$. Indeed, $b$ has been expressed as a function of $m$ and differential equation for $a$ and $b$ are function of $(b, P, \phi)$. Then, we end up with the following system of first order differential equation
\bea
\label{eq:systemTOV}
    &&\dot{P}= -\dfrac{F_1}{2}(P+\rho)+\dfrac{\Psi}{2\phi}(\mathcal{L}_m-P) \\
    &&\dot{m}=4\pi r^2\left(\frac{\rho}{\sqrt{\phi}}-\frac{F_3}{\Tilde{\kappa}}\right) \\
    &&\dot{\phi}=\Psi \\
    &&\dot{\Psi}=-\dfrac{\Psi}{2}\left[F_1-F_2+\dfrac{4}{r} \right]+\dfrac{\Tilde{\kappa}\sqrt{\phi}}{3}b(T-\mathcal{L}_m)\label{eq:systemTOVe}
\eea
with
\bea
    && F_1 =\dfrac{b}{r}\left(\dfrac{Pr^2\Tilde{\kappa}}{\sqrt{\phi}}+1-\dfrac{1}{b}-\dfrac{2r}{b}\dfrac{\Psi}{\phi}\right)\left(1+\dfrac{r}{2}\dfrac{\Psi}{\phi}\right)^{-1} \\
    && F_2 =\dfrac{1-b}{r}+rb\left(\dfrac{\Tilde{\kappa} \rho}{\sqrt{\phi}}-F_3\right) \\
    && F_3 =\dfrac{1}{2b}\dfrac{\Psi}{\phi}F_1+\dfrac{\Tilde{\kappa}}{3\sqrt{\phi}}(\Lm-T) \\
    && b = \dfrac{1}{1-m\Tilde{\kappa}/4\pi r}
\eea
\section{Numerical implementation and results}

The code, as well as a script that generates all the figures that are presented in this manuscript, are freely available on github \cite{code}.
\subsection{Equation of state}

We assume a simple polytropic equation of state in this exploratory study, as the goal of this study is to see the broad behavior of compact object solutions in entangled relativity, compared to general relativity. Besides, the equations for matter fields are modified in entangled relativity due to the non-minimal coupling between geometry and matter---see Eq. (\ref{eq:noncons}). As a consequence, nuclear physics may be impacted at high density, which leads to many new open questions that are beyond the scope of this manuscript.

The equation of state that we shall consider reads
\be \label{eq:EqS}
P=K \rho^{\gamma},
\ee
with $\gamma = 5/3$ and $K = 1.475 \times 10^{-3} (fm^3/MeV)^{2/3}$ \cite{schlogel:2014pr}.

We shall limit our investigations to solutions that have a sub-luminal speed of sound $v_S$ everywhere in the compact object. The condition implemented in our code reads
\be
\frac{v_S}{c} \equiv  \left(\frac{d P}{d \rho} \right)^{1/2} =  \left(\gamma \frac{P}{\rho} \right)^{1/2} < 1~ \forall r \in ~]0,R_*],
\ee
where $R_*$ is the radius of the compact object.

\subsection{On-shell matter Lagrangian}

Now, one ought to know what is the value of the on-shell matter Lagrangian.\footnote{By \textit{on-shell}, here we mean the effective value that takes the Lagrangian when the matter field solutions are injected into its formal definition.} Eventually, one may be able to derive it from first principles. However, the task is somewhat complex, given the additional complication that the matter field equations are modified with respect to general relativity, due to the non-minimal coupling between matter and geometry in entangled relativity---for instance, the modified Maxwell equation reads $\nabla_{\nu}\left(\sqrt{\phi} F^{\mu \nu}\right)=0$, where $F^{\mu \nu}$ is the usual electromagnetic tensor.

In what follows we will assume an effective perfect fluid description of matter fields, whose matter field density is conserved $\nabla_\sigma (\rho_0 u^\sigma) = 0$---where $\rho_0$ is the rest-mass energy density---such that one can show that the matter Lagrangian reduces to $\mathcal{L}_m = -\rho$, where $\rho$ is the total energy density of the fluid \cite{minazzoli:2012pr,minazzoli:2013pd}---see Appendix \ref{app:cons}. In that case, one can see in Eq. (\ref{eq:sceq}) that the scalar-field would only be sourced by pressure, such that this type of scalar-fields has been dubbed \textit{pressuron} in \cite{minazzoli:2014pr}.

However, it has recently been argued that for generic solitons, the on-shell Lagrangian ought to reduce to the trace of the stress-energy tensor $\mathcal{L}_m = T$ \cite{avelino:2018pr,*avelino:2018pd}. If that argument held for compact objects such as neutron stars, one can see from Eq. (\ref{eq:sceq}) that the decoupling of the scalar-field would be total, such that one could expect to recover the results of general relativity. We checked that if we assume $\mathcal{L}_m = T$ instead of $\mathcal{L}_m = -\rho$, we indeed recover the TOV solutions of general relativity. \footnote{Note that entangled relativity would nevertheless be different from general relativity for any matter field that do not satisfy $\mathcal{L}_m = T$, such as, for instance a pure magnetic field--- which leads to $\mathcal{L}_m \propto B^2 \neq T$ \cite{thorsrud:2012jh}---or an homogeneous scalar field---which leads to $\mathcal{L}_m = P \neq T$ \cite{turner:1983pr}. In particular, one can expect the Higgs field to lead to a different behavior of the cosmic solution of entangled relativity close to the big bang, compared to general relativity.}

Another interesting thing to keep in mind is that, since the on-shell Lagrangian $\mathcal{L}_m$ depends on the field solutions, one cannot a priori exclude the possibility that it could transition from, say, $\mathcal{L}_m=-\rho$ to $\mathcal{L}_m=T$, or $\mathcal{L}_m=T$ to $\mathcal{L}_m=P$, etc., after a phase transition of matter fields at some density inside the compact object. Studying this is beyond the scope of the present paper, as it would require to derive the behavior of matter fields from first principles; whereas we assume here an effective perfect fluid description. What is clear however, is that in the dust limit, one has $\mathcal{L}_m = -\rho = T$. Hence, $\mathcal{L}_m = P$ seems to be excluded, unless the aforementioned transition of phase can occur at high enough energy or pressure. But again, this is beyond the scope of the present manuscript.
\subsection{Scalar field normalisation}

We normalize the scalar field such that $\tilde \kappa / \sqrt{\phi_L} = 8 \pi G / c^4$, where $G$ is the measured value of the constant of Newton in the solar system, and $\phi_L$ the asymptotic value of the scalar field at a distance $r=L$ where the derivative of the gravitational potential becomes negligible, such that it matches the zeroth order of the post-Newtonian expansion of the theory in the weak-field regime of the solar system \cite{minazzoli:2013pr}. Note however that in practice, the effective value of $\phi_L$ can in principle take different values at different locations of the universe, which corresponds in effect to a constant of Newton that actually depends on the location. However, given the fact that the scalar field is not sourced by pressureless matter fields in the weak field regime \cite{minazzoli:2013pr}, one does not expect the two values to differ significantly in a dust\footnote{Baryonic and cold dark matter.} dominated environment such as our own galaxy.

Numerically, the choice $\tilde \kappa=\kappa$ has been made, so that the normalization of the scalar field reads $\phi_\infty\equiv\phi_L=1$. The goal is then to find the central value $\phi_0$ that leads to that normalization. In order to do so, one could run the simulation several times with different central values of the scalar field, until the one that corresponds to the desired normalization is reached. However, for the sake of computational speed, one could take the advantage of the following property: 
if $(\phi,\psi,m,P)(r)$ are solution of (\ref{eq:systemTOV}-\ref{eq:systemTOVe}), then $(\alpha\phi,\psi,\alpha^{-1/4}m,\alpha P)(\alpha^{-1/4}r)$ are another solution $\forall\alpha$. If a simulation starts with a random initial value $\phi_0$ and lead to a star radius $\mathfrak{R}_*$, and an asymptotic scalar field value $\phi_\infty$, then, the radius with the appropriate units is $R_*=\phi_\infty^{1/4}\mathfrak{R}_*$, where the property has been used with $\alpha=\phi_\infty^{-1}$.

\subsection{Mass and radius comparison}

\begin{figure}
\includegraphics[scale=0.6]{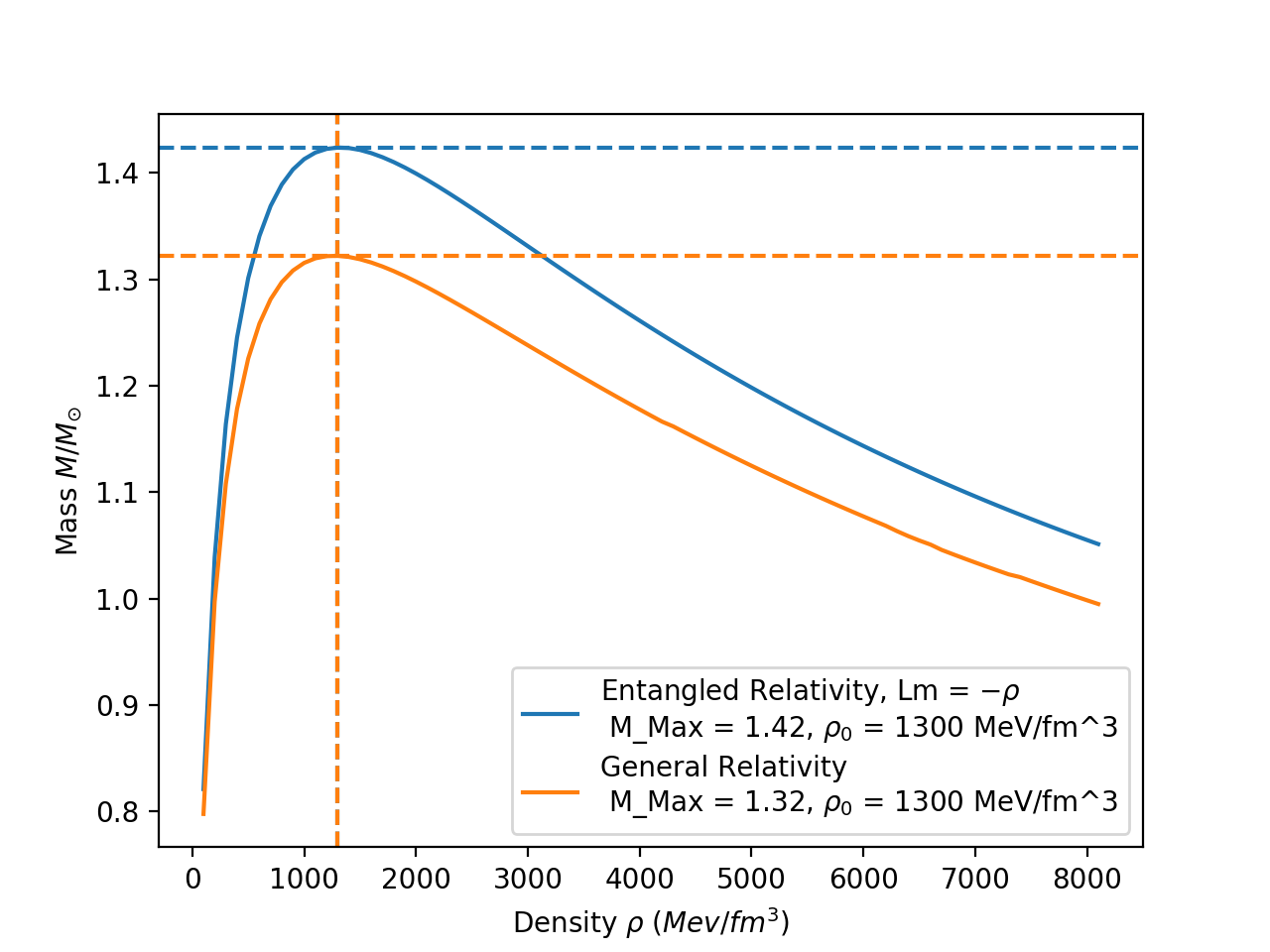}
\caption{Diagram of the mass $m_*$ of the compact object with respect to its central density, for central densities between 100 and 8200 $Mev/fm^3$.}\label{fig:Mvsrho}
\end{figure}

\begin{figure}
\includegraphics[scale=0.6]{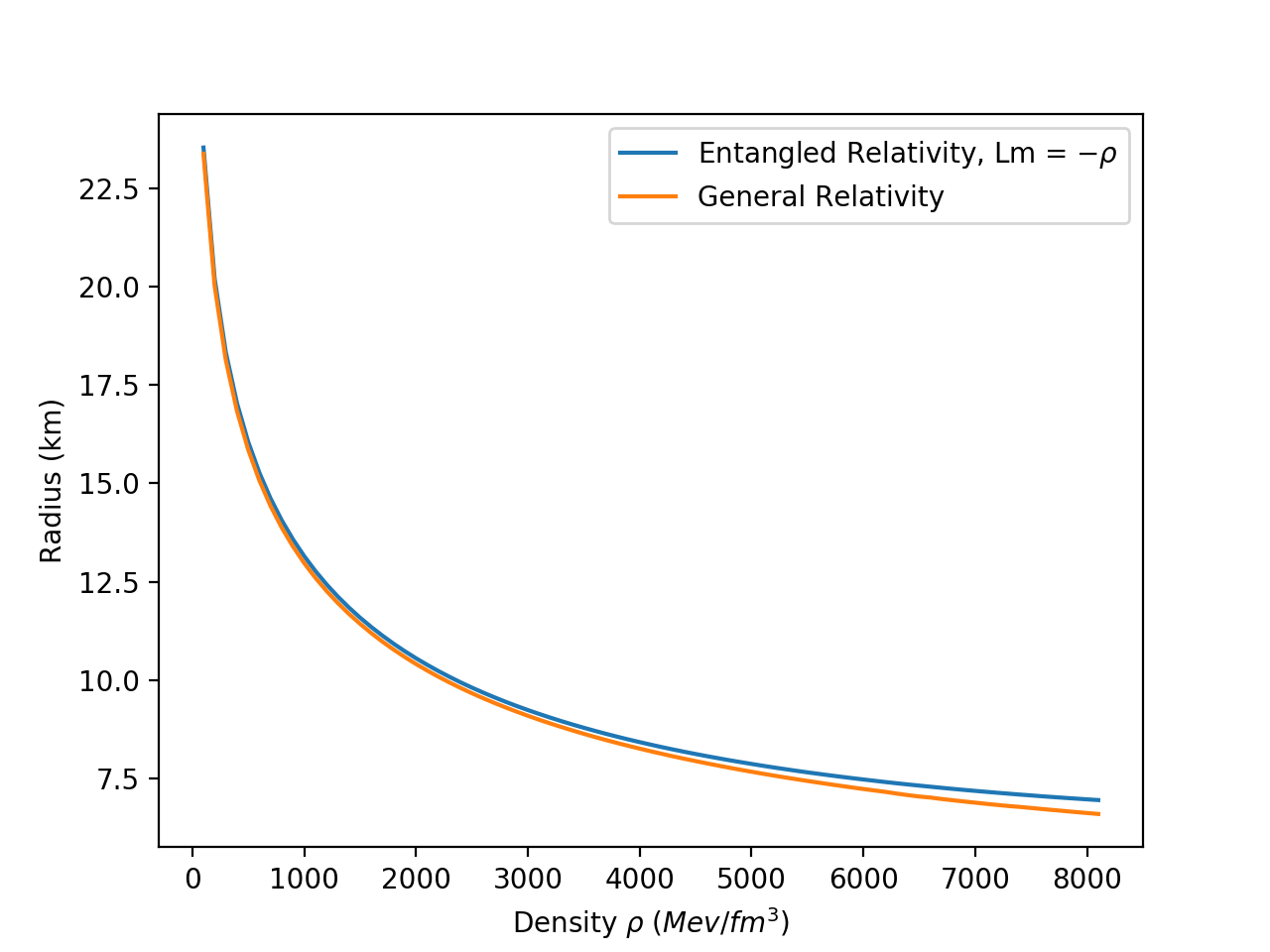}
\caption{Diagram of the radius of the compact object with respect to its central density, for central densities between 100 and 8200 $Mev/fm^3$.}\label{fig:Rvsrho}
\end{figure}

Because the dilaton field is sourced by pressure when $\mathcal{L}_m = -\rho$, and because the pressure increases with the density, one expects to have an increasing deviation from general relativity with the growth of the density. In the figures \ref{fig:Mvsrho} and \ref{fig:Rvsrho}, one can see that this is broadly the case, although one should have a look at the metric in different cases in order to have a full picture of this. In particular, the denser the object, the more the metric deviates from the Schwarzschild metric outside the compact object---see Sec. \ref{sec:metrics}. Also, one can see that, for central densities between 100 and 8200 $Mev/fm^3$, both the mass and the radius of the compact object are always higher in entangled relativity. The limit at  8200 $Mev/fm^3$ is set by the condition that the speed of sound in the compact object never exceeds the causality speed limit $c$.

\begin{figure}
\includegraphics[scale=0.6]{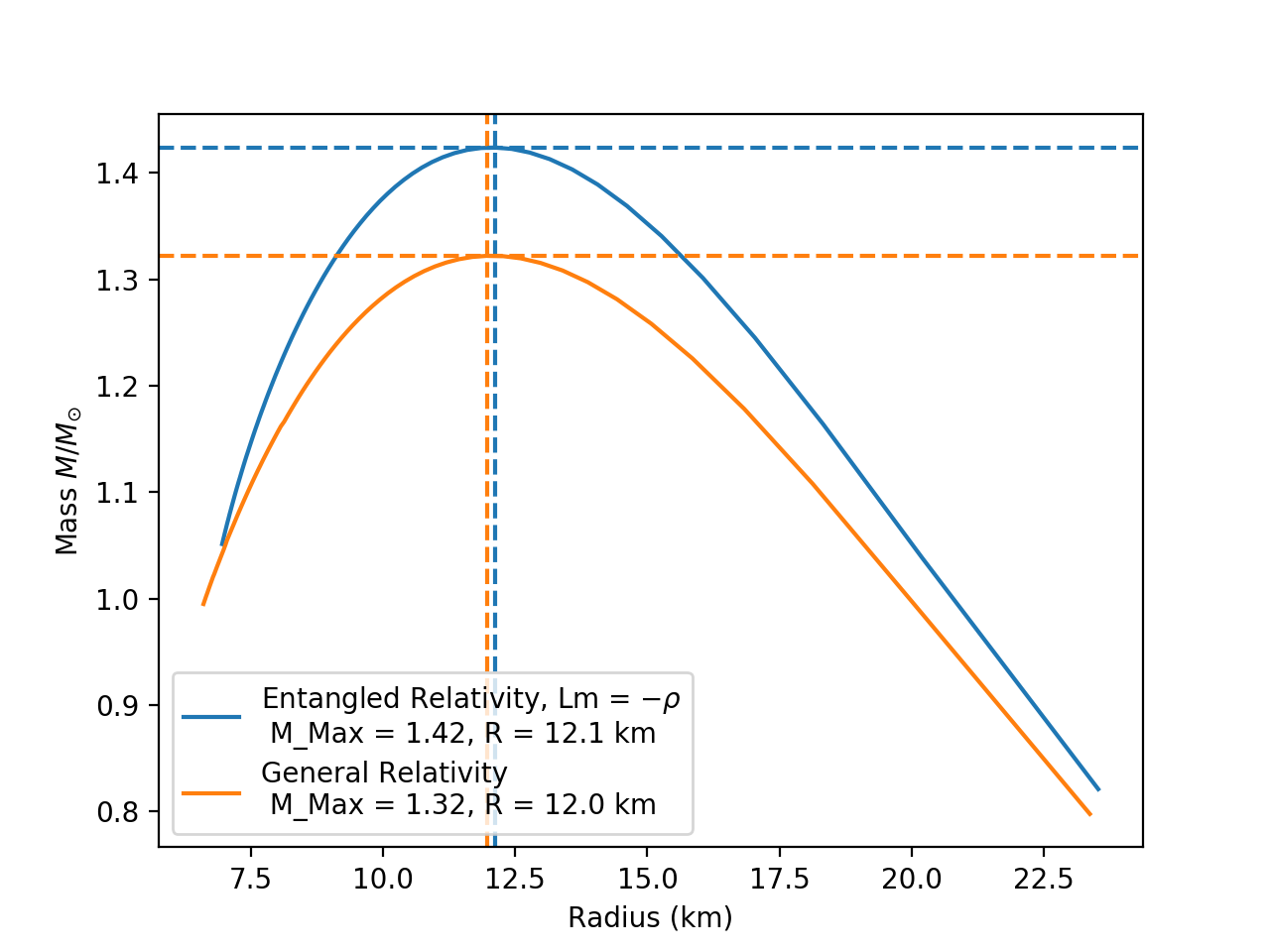}
\caption{Diagram of the mass $m_*$ of the compact object with respect to its radius in both theories, for central densities between 100 and 8200 $Mev/fm^3$.}\label{fig:MvsR}
\end{figure}

\begin{figure}
\includegraphics[scale=0.6]{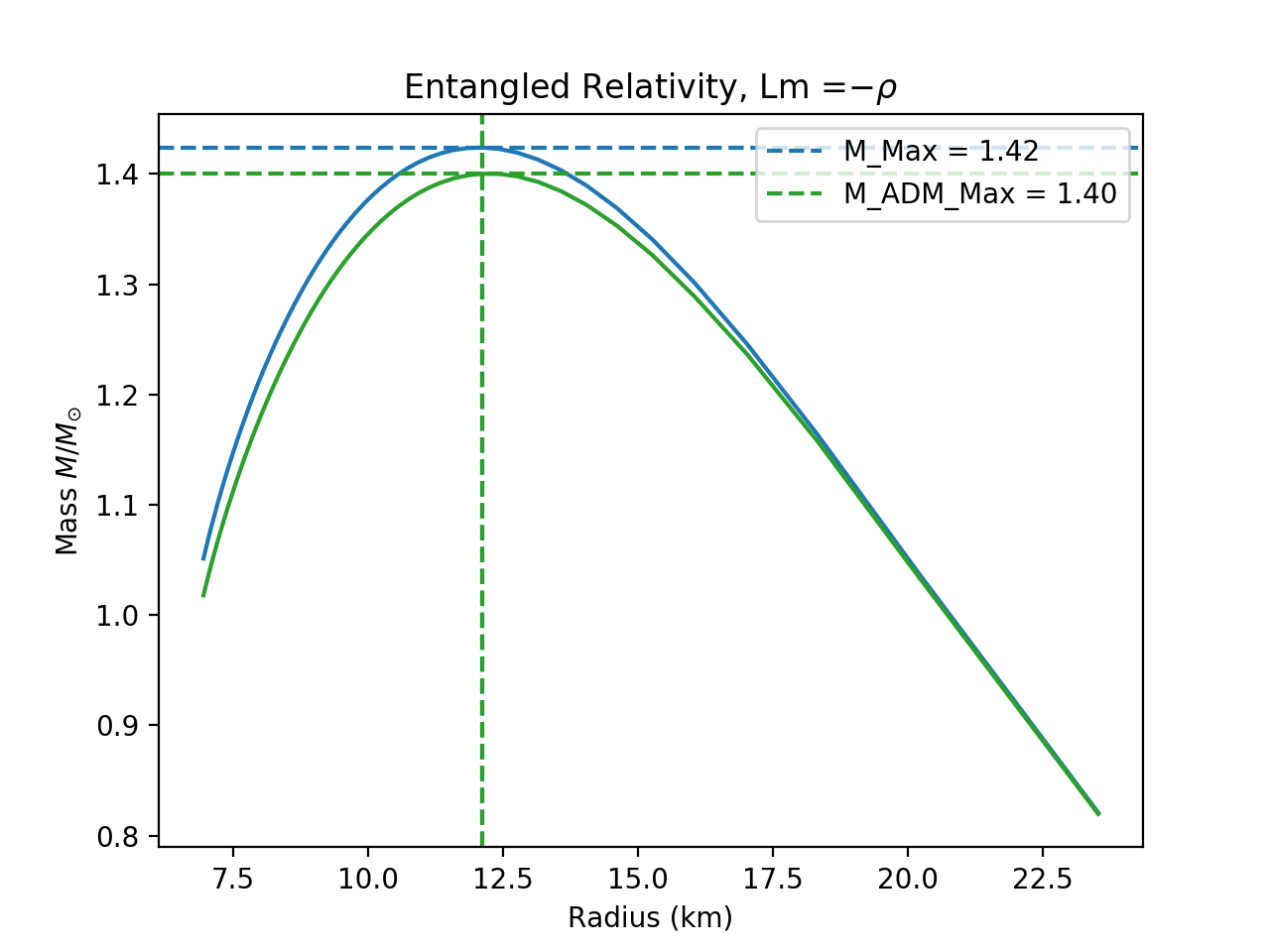}
\caption{Diagram of the masses $m_*$ and $\mathfrak{M}_{ADM}$ of the compact object with respect to its radius in entangled relativity, for central densities between 100 and 8200 $Mev/fm^3$.}\label{fig:adm}
\end{figure}

One can see that, for a given central density, the mass $m_*$---defined in Eq. (\ref{eq:massatrad})---is always bigger in entangled relativity with respect to general relativity. The same is true for the ADM mass $\mathfrak{M}_{ADM}$---defined in Eq. (\ref{eq:mADM})---, although the ADM mass is always smaller than the mass $m_*$, as one can see in Fig. \ref{fig:adm}. The difference between $m_*$ and $\mathfrak{M}_{ADM}$ simply comes from the infinite range contribution of the scalar-field to the mass---defined in Eq. (\ref{eq:m}). 

At this stage, it would be tempting to argue that, as a consequence, entangled relativity may be a good candidate to explain the relatively high masses of neutron stars observed over the last decade \cite{alsing:2018mn}, and which seem to lead to a tension between astrophysical observations and the union of general relativity and nuclear physics \cite{biswas:2020ar}. However, this is way too premature given that one has to check first how the observables that are used to infer those masses from observations are actually affected in entangled relativity. For instance, one has to see how the Shapiro delay is modified in entangled relativity with respect to general relativity, and see how the inferred mass depends on the adjustment of the data in the new framework of entangled relativity. One has to note in particular that, unlike in general relativity, the mass of a compact object---defined in Eq. (\ref{eq:m})---continues to change outside the compact object due to the contribution of the scalar-field. It continuously goes from $m_*$---defined in Eq. (\ref{eq:massatrad})---to $\mathfrak{M}_{ADM}$---defined in Eq. (\ref{eq:mADM}). This will affect observables that are sensitive to the metric outside the compact object---see Sec. \ref{sec:howto}.

Another interesting thing to point out is that, for a given central density, there is not a big difference in terms of radius between the predictions of general relativity and entangled relativity, as one can see in Fig. \ref{fig:Rvsrho}.
 On the other hand, for a given mass, the radius of high density (small radius) objects in entangled relativity are always smaller than in general relativity; whereas it is the opposite for low density (large radius) objects \footnote{resp. l.h.s. and r.h.s. of Fig. \ref{fig:MvsR}.}. 

\subsection{Scalar field behavior}

The scalar field grows monotonically from the center to infinity---where it reaches its normalized value $\phi_\infty = 1$. In both the cases that lead to, e.g., $M=1.25 M_\odot$---that is, the solutions with $400 ~MeV/fm^3$ and $4150 ~MeV/fm^3$ central densities---the scalar field derivative reaches a maximum toward roughly half the radius before monotonically decreasing to zero. Since matter physics depends on this derivative---see for instance Eq. (\ref{eq:noncons})---it means that most of the effect of the scalar field happens inside the compact object. Nevertheless, the scalar field derivative is not null outside the compact object, such that one can expect entangled relativity to have a richier phenomenology with respect to general relativity---because of the scalar field's behavior outside the compact object.

\begin{figure}
\includegraphics[scale=0.5]{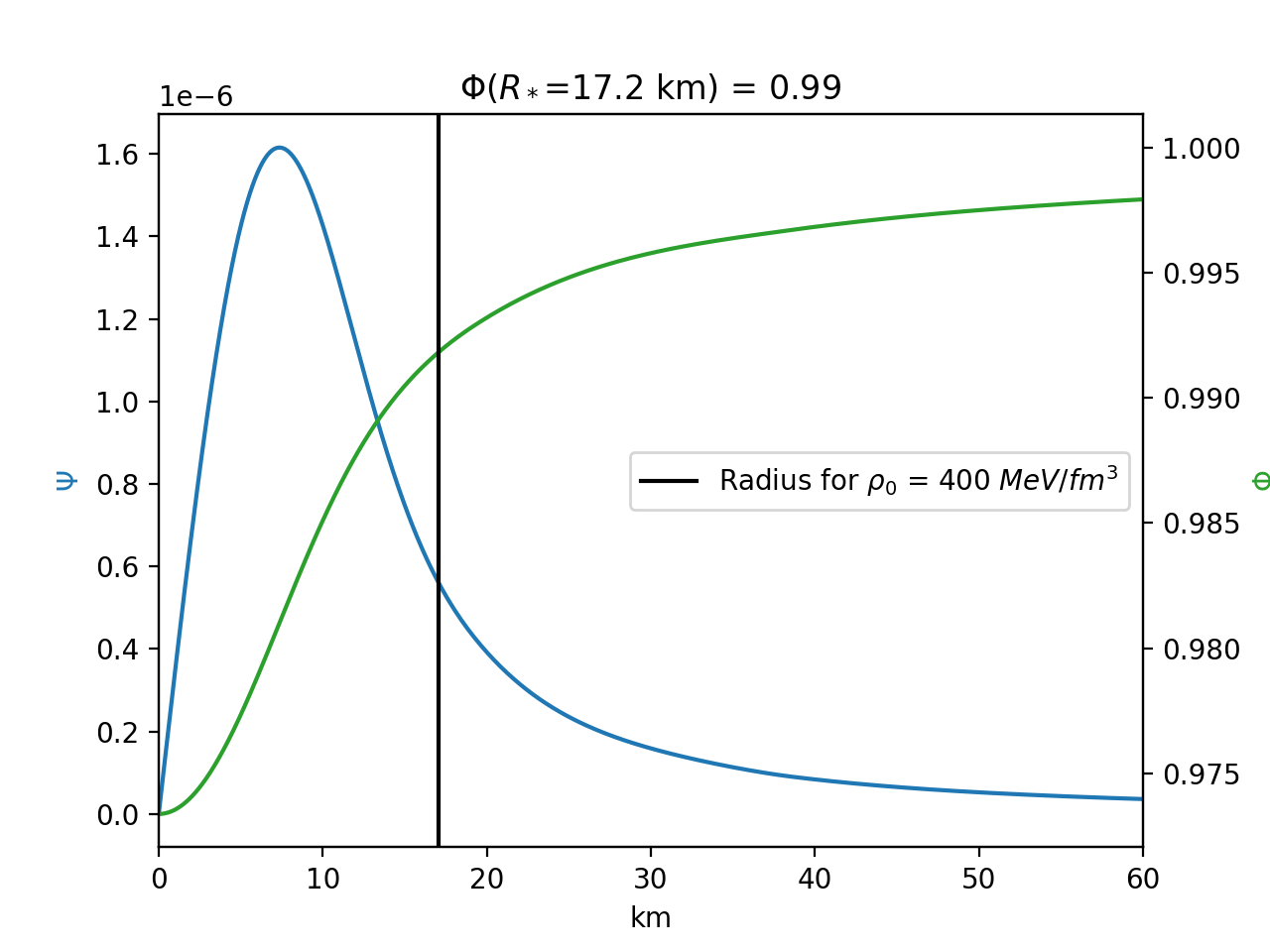}
\includegraphics[scale=0.5]{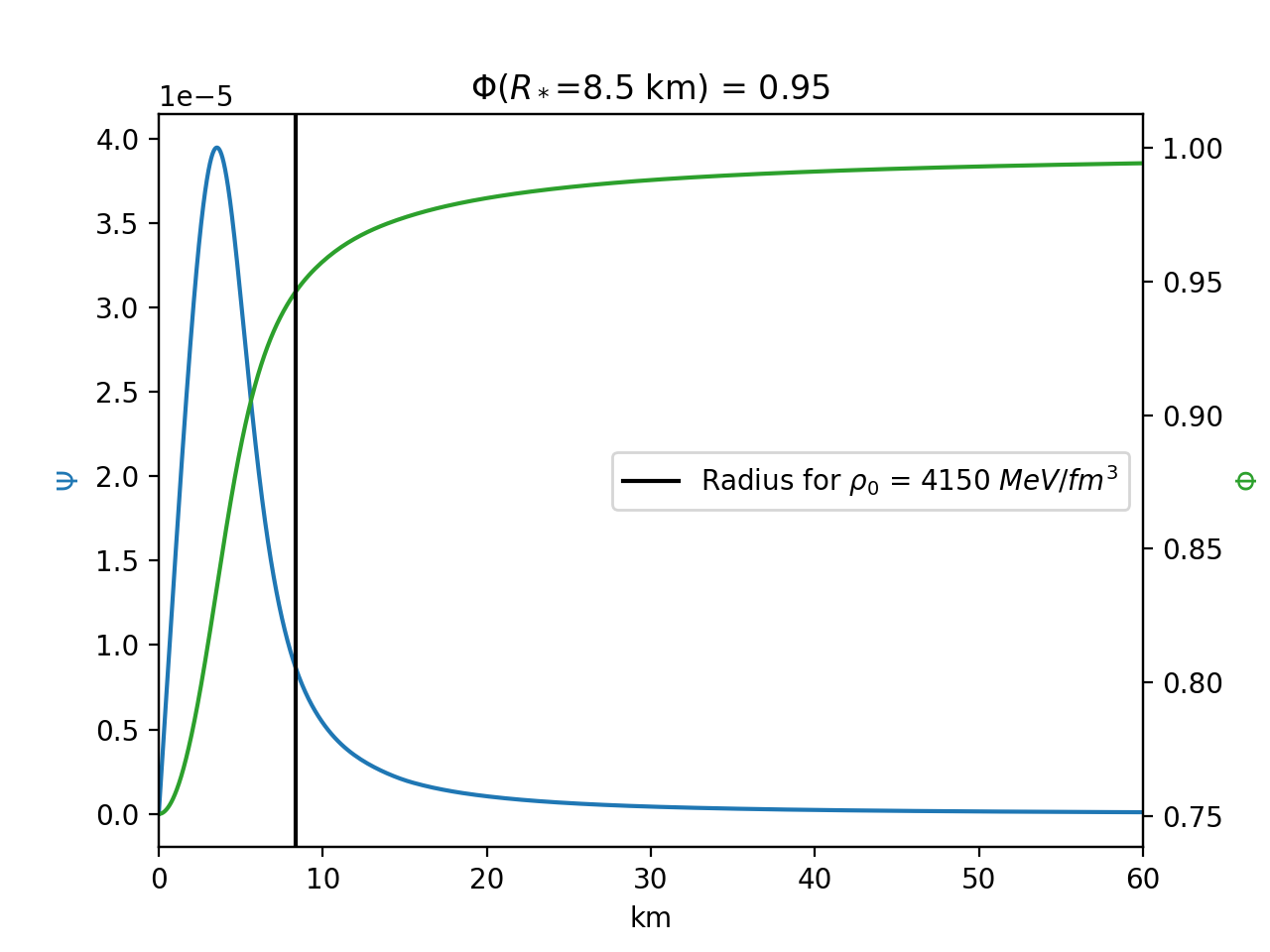}
\caption{Scalar field and its derivative profiles for the two solutions in entangled relativity with $M=1.25 M_\odot$, with densities $400 ~MeV/fm^3$ or $4150 ~MeV/fm^3$ ($R_*=17.0$ or $R_*=8.3$ km respectively). The vertical lines mark the radius of the solutions.}\label{fig:phi_psi}
\end{figure}

\subsection{Metrics comparison}
\label{sec:metrics}

\begin{figure}
\includegraphics[scale=0.6]{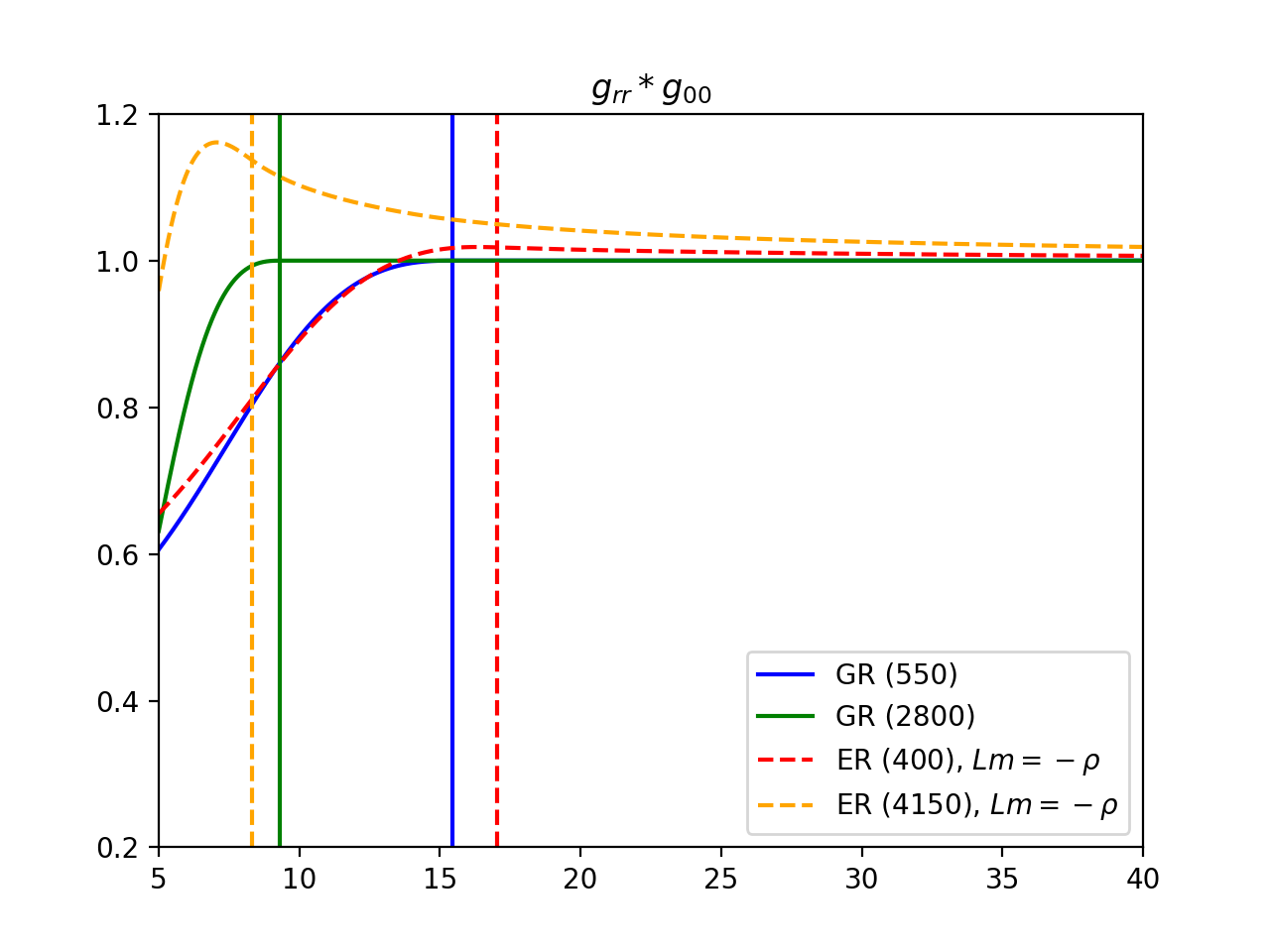}
\caption{Product of the metric's components $g_{00} g_{rr}$ of the two solutions of an object of mass $M = 1.25 M_\odot$ in general relativity and entangled relativity. The vertical lines mark the radius of the solutions. In the parenthesis are displayed the central density for each solution, in $MeV/fm^3$.}\label{fig:g00grr}
\end{figure}

Most of the time, the mass of the compact object is degenerate with respect to the value of its central density, as one can see in Fig. \ref{fig:Mvsrho}. In general relativity, the specific value of the central density for two solutions with the same mass only plays a role with respect to the radius of the compact object, as Birkhoff's theorem implies that the external metric ought to be the same for different solutions of an object of a given mass. However, Birkhoff's theorem no longer holds in entangled relativity. Notably, unlike the case of general relativity, the scalar product between the time-time and space-space components of the metric outside a body is not a constant in entangled relativity. In Fig. \ref{fig:g00grr}, we show the product of the metric's components $g_{00} g_{rr}$ of the two solutions of an object of mass $M = 1.25 M_\odot$ for each theory. In general relativity, the central density for this mass $M$ can either be around $550 ~MeV/fm^3$ or $2800 ~MeV/fm^3$, corresponding to a radius of 15.4 km or 9.3 km respectively; whereas in entangled relativity, it can either be around $400 ~MeV/fm^3$ or $4150 ~MeV/fm^3$, corresponding to a radius of 17.0 km or 8.3 km respectively. 

As one can see, one has $g_{00} g_{rr} =$ cst outside the compact object in general relativity in each case, whereas it is no longer the case in entangled relativity. In particular, the denser the object, the more different is the metric's components with respect to general relativity. It is the expected behavior, given that the scalar field is more \textit{sourced} for denser objects.

\begin{figure}
\includegraphics[scale=0.6]{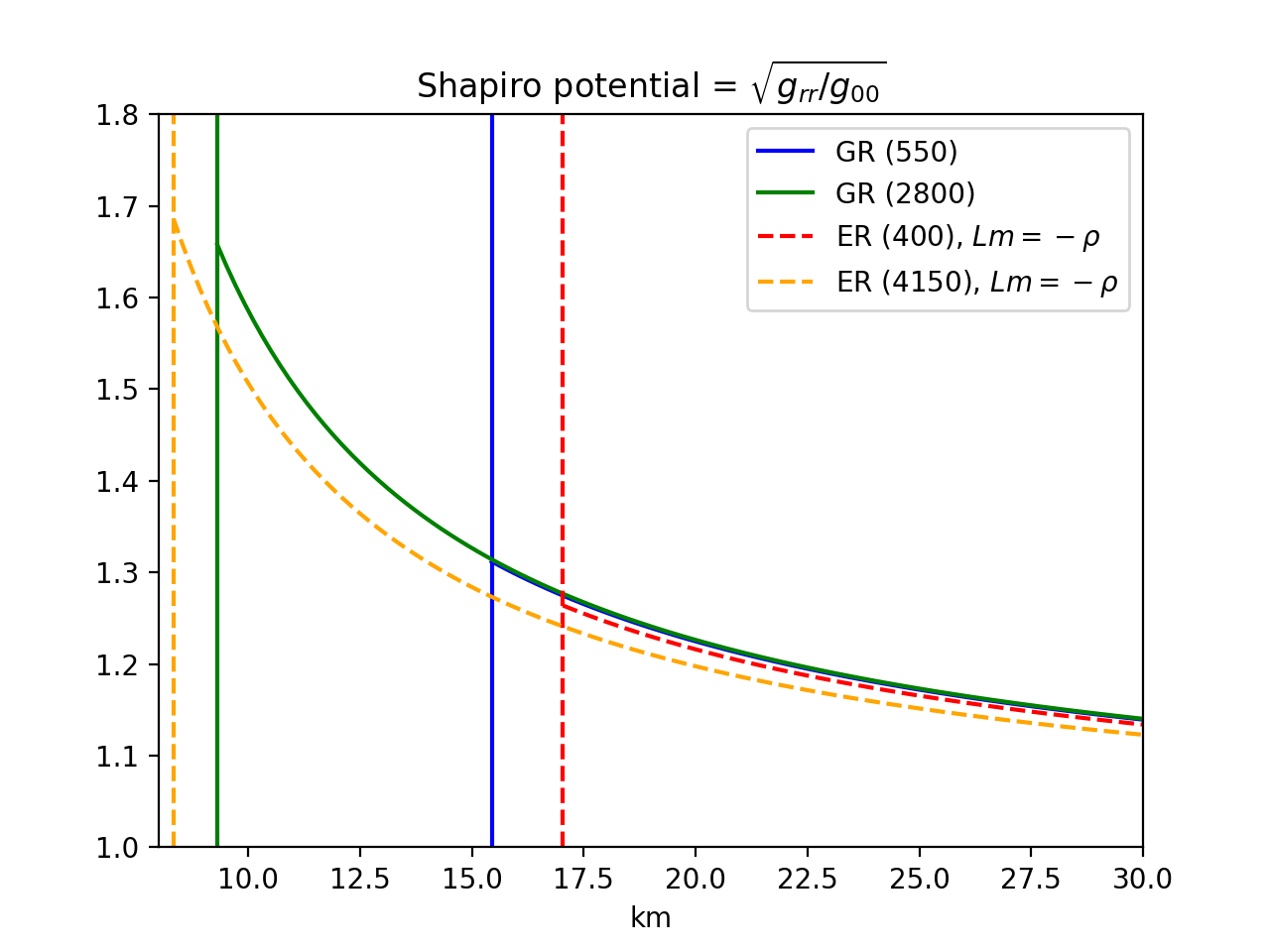}
\caption{Shapiro potential with respect to the radius for each solution of mass $M = 1.25 M_\odot$  in both theories. The vertical lines mark the radius of the solutions. In the parenthesis are displayed the central density for each solution, in $MeV/fm^3$.}\label{fig:shapiro_pot}
\end{figure}

Since the metric's components deviate from the one of general relativity, one should expect the Shapiro delay to be modified accordingly. In particular, we shall expect that (what we shall call) the \textit{Shapiro potential}---defined here as equal to $\sqrt{g_{rr}/g_{00}}$---should be modified with respect to general relativity. However, we see in Fig. \ref{fig:shapiro_pot} that the Shapiro potential remains qualitatively the same in entangled relativity.

\subsection{Discussion on how to test the theory}
\label{sec:howto}

The prediction of the mass and radius of neutron stars is not enough in order to test the theory for two reasons. First, those quantities depend significantly on the unknown equation of state of neutron stars. As a consequence, it may always be possible to tune the equation of state to recover the observed radius and mass. Second, the mass is not an observable, but is inferred from observables---such as the differential Shapiro time delay. Hence, one has to readjust the model parameters (such as the mass and the radius) from observations within the new framework before being able to say whether or not the theory explains the actual data well. For instance, regarding the Shapiro delay, because the metric outside the neutron star is not the Schwarzschild metric, the differential Shapiro delay does not follow the equation of general relativity that links the effect to the mass of the spherical compact object. Therefore, the mass of various neutron stars that have been inferred by measuring the differential Shapiro delay and assuming general relativity \cite{demorest:2010na,fonseca:2016aj,cromartie:2020na} are likely no longer the best fit for entangled relativity. Unfortunately, because the Birkhoff's theorem is not valid in entangled relativity, the differential Shapiro delay is not uniquely defined with the mass of the source and the differential distance of the light-ray to the source. For instance, one can see in Fig. \ref{fig:g00grr} that the outside metrics for two more or less dense compact objects with the same mass are not the same in entangled relativity---unlike what happens in general relativity. This may complicate the inversion from the observables to the model parameters in entangled relativity.

A potential way to test the theory would be to use X-ray pulse profiles from NASA's Neutron star Interior Composition Explorer (NICER) \cite{gendreau:2017na}, as it was recently proposed in order to probe strong-field effects in scalar-tensor theory \cite{silva:2019pr,silva:2019cq,xu:2020pr}. In order to do so, one would have to extend the present work to the rotating case, as well as to consider more realistic equations of state. However, let us stress again that the matter field equations are modified in entangled relativity with respect to general relativity. This may induce yet another complication in such a study. Finally, let us note that unlike in the cases considered in \cite{silva:2019pr,silva:2019cq,xu:2020pr}, the electromagnetic intensity is not conserved in entangled relativity due to the non-minimal coupling between the scalar degree of freedom and the electromagnetic field in the action \cite{minazzoli:2014pr, *minazzoli:2014pl}. This additional effect will have to be taken into account in order to correctly predict the variety of X-ray pulse profiles in entangled relativity.

Obviously, another way to test the theory would be through the observation of gravitational waves that have been emitted during the fusion of neutron stars \cite{abbott:2017pl}---although, again, the potential degeneracy with the unknown neutron star equation of state might limit the constraining power of such tests. This has yet to be explored.

\section{Conclusion}

We provided the first study of compact objects within the framework of entangled relativity.  Assuming a simple polytropic equation of state and the conservation of the rest-mass density, we showed that, for any given density, compact objects are always heavier in entangled relativity than in general relativity---for any given central density within the usual range of neutron stars central densities, or for a given radius of the resulting compact object. Because entangled relativity is parameter free at the classical level, the results presented in the manuscript do not depend on any parameter. Nevertheless, they are still dependent on a perfect fluid description of matter fluid, and on the assumption that the rest-mass density is conserved---which is not a straightforward property in theories with a non-minimal coupling between a scalar degree of freedom and matter fields, see Appendix \ref{app:cons}. This would eventually have to be obtained from first principles with an accurate description of the microphysical properties of nuclear matter in neutron stars. But this may turn out to be very demanding, notably because of the modifications of the usual matter field equations due to the non-minimal scalar-matter coupling in the action Eq. (\ref{eq:sfaction}).

\bibliography{TOV_ER}

\begin{thebibliography}{27}%
\makeatletter
\providecommand \@ifxundefined [1]{%
 \@ifx{#1\undefined}
}%
\providecommand \@ifnum [1]{%
 \ifnum #1\expandafter \@firstoftwo
 \else \expandafter \@secondoftwo
 \fi
}%
\providecommand \@ifx [1]{%
 \ifx #1\expandafter \@firstoftwo
 \else \expandafter \@secondoftwo
 \fi
}%
\providecommand \natexlab [1]{#1}%
\providecommand \enquote  [1]{``#1''}%
\providecommand \bibnamefont  [1]{#1}%
\providecommand \bibfnamefont [1]{#1}%
\providecommand \citenamefont [1]{#1}%
\providecommand \href@noop [0]{\@secondoftwo}%
\providecommand \href [0]{\begingroup \@sanitize@url \@href}%
\providecommand \@href[1]{\@@startlink{#1}\@@href}%
\providecommand \@@href[1]{\endgroup#1\@@endlink}%
\providecommand \@sanitize@url [0]{\catcode `\\12\catcode `\$12\catcode
  `\&12\catcode `\#12\catcode `\^12\catcode `\_12\catcode `\%12\relax}%
\providecommand \@@startlink[1]{}%
\providecommand \@@endlink[0]{}%
\providecommand \url  [0]{\begingroup\@sanitize@url \@url }%
\providecommand \@url [1]{\endgroup\@href {#1}{\urlprefix }}%
\providecommand \urlprefix  [0]{URL }%
\providecommand \Eprint [0]{\href }%
\providecommand \doibase [0]{http://dx.doi.org/}%
\providecommand \selectlanguage [0]{\@gobble}%
\providecommand \bibinfo  [0]{\@secondoftwo}%
\providecommand \bibfield  [0]{\@secondoftwo}%
\providecommand \translation [1]{[#1]}%
\providecommand \BibitemOpen [0]{}%
\providecommand \bibitemStop [0]{}%
\providecommand \bibitemNoStop [0]{.\EOS\space}%
\providecommand \EOS [0]{\spacefactor3000\relax}%
\providecommand \BibitemShut  [1]{\csname bibitem#1\endcsname}%
\let\auto@bib@innerbib\@empty
\bibitem [{\citenamefont {{Ludwig}}\ \emph {et~al.}(2015)\citenamefont
  {{Ludwig}}, \citenamefont {{Minazzoli}},\ and\ \citenamefont
  {{Capozziello}}}]{ludwig:2015pl}%
  \BibitemOpen
  \bibfield  {author} {\bibinfo {author} {\bibfnamefont {Hendrik}\ \bibnamefont
  {{Ludwig}}}, \bibinfo {author} {\bibfnamefont {Olivier}\ \bibnamefont
  {{Minazzoli}}}, \ and\ \bibinfo {author} {\bibfnamefont {Salvatore}\
  \bibnamefont {{Capozziello}}},\ }\bibfield  {title} {\enquote {\bibinfo
  {title} {{Merging matter and geometry in the same Lagrangian}},}\ }\href
  {\doibase 10.1016/j.physletb.2015.11.023} {\bibfield  {journal} {\bibinfo
  {journal} {Physics Letters B}\ }\textbf {\bibinfo {volume} {751}},\ \bibinfo
  {pages} {576--578} (\bibinfo {year} {2015})},\ \Eprint
  {http://arxiv.org/abs/1506.03278} {arXiv:1506.03278 [gr-qc]} \BibitemShut
  {NoStop}%
\bibitem [{\citenamefont {{Minazzoli}}(2018)}]{minazzoli:2018pr}%
  \BibitemOpen
  \bibfield  {author} {\bibinfo {author} {\bibfnamefont {Olivier}\ \bibnamefont
  {{Minazzoli}}},\ }\bibfield  {title} {\enquote {\bibinfo {title} {{Rethinking
  the link between matter and geometry}},}\ }\href {\doibase
  10.1103/PhysRevD.98.124020} {\bibfield  {journal} {\bibinfo  {journal}
  {\prd}\ }\textbf {\bibinfo {volume} {98}},\ \bibinfo {eid} {124020} (\bibinfo
  {year} {2018})},\ \Eprint {http://arxiv.org/abs/1811.05845} {arXiv:1811.05845
  [gr-qc]} \BibitemShut {NoStop}%
\bibitem [{\citenamefont {{Minazzoli}}\ and\ \citenamefont
  {{Hees}}(2013)}]{minazzoli:2013pr}%
  \BibitemOpen
  \bibfield  {author} {\bibinfo {author} {\bibfnamefont {Olivier}\ \bibnamefont
  {{Minazzoli}}}\ and\ \bibinfo {author} {\bibfnamefont {Aur{\'e}lien}\
  \bibnamefont {{Hees}}},\ }\bibfield  {title} {\enquote {\bibinfo {title}
  {{Intrinsic Solar System decoupling of a scalar-tensor theory with a
  universal coupling between the scalar field and the matter Lagrangian}},}\
  }\href {\doibase 10.1103/PhysRevD.88.041504} {\bibfield  {journal} {\bibinfo
  {journal} {\prd}\ }\textbf {\bibinfo {volume} {88}},\ \bibinfo {eid} {041504}
  (\bibinfo {year} {2013})},\ \Eprint {http://arxiv.org/abs/1308.2770}
  {arXiv:1308.2770 [gr-qc]} \BibitemShut {NoStop}%
\bibitem [{\citenamefont {{Minazzoli}}\ and\ \citenamefont
  {{Hees}}(2014)}]{minazzoli:2014pr}%
  \BibitemOpen
  \bibfield  {author} {\bibinfo {author} {\bibfnamefont {Olivier}\ \bibnamefont
  {{Minazzoli}}}\ and\ \bibinfo {author} {\bibfnamefont {Aur{\'e}lien}\
  \bibnamefont {{Hees}}},\ }\bibfield  {title} {\enquote {\bibinfo {title}
  {{Late-time cosmology of a scalar-tensor theory with a universal
  multiplicative coupling between the scalar field and the matter
  Lagrangian}},}\ }\href {\doibase 10.1103/PhysRevD.90.023017} {\bibfield
  {journal} {\bibinfo  {journal} {\prd}\ }\textbf {\bibinfo {volume} {90}},\
  \bibinfo {eid} {023017} (\bibinfo {year} {2014})},\ \Eprint
  {http://arxiv.org/abs/1404.4266} {arXiv:1404.4266 [gr-qc]} \BibitemShut
  {NoStop}%
\bibitem [{\citenamefont {{Minazzoli}}(2014)}]{minazzoli:2014pl}%
  \BibitemOpen
  \bibfield  {author} {\bibinfo {author} {\bibfnamefont {Olivier}\ \bibnamefont
  {{Minazzoli}}},\ }\bibfield  {title} {\enquote {\bibinfo {title} {{On the
  cosmic convergence mechanism of the massless dilaton}},}\ }\href {\doibase
  10.1016/j.physletb.2014.06.027} {\bibfield  {journal} {\bibinfo  {journal}
  {Physics Letters B}\ }\textbf {\bibinfo {volume} {735}},\ \bibinfo {pages}
  {119--121} (\bibinfo {year} {2014})},\ \Eprint
  {http://arxiv.org/abs/1312.4357} {arXiv:1312.4357 [gr-qc]} \BibitemShut
  {NoStop}%
\bibitem [{\citenamefont {{Minazzoli}}\ and\ \citenamefont
  {{Hees}}(2016)}]{minazzoli:2016pr}%
  \BibitemOpen
  \bibfield  {author} {\bibinfo {author} {\bibfnamefont {Olivier}\ \bibnamefont
  {{Minazzoli}}}\ and\ \bibinfo {author} {\bibfnamefont {Aur{\'e}lien}\
  \bibnamefont {{Hees}}},\ }\bibfield  {title} {\enquote {\bibinfo {title}
  {{Dilatons with intrinsic decouplings}},}\ }\href {\doibase
  10.1103/PhysRevD.94.064038} {\bibfield  {journal} {\bibinfo  {journal}
  {\prd}\ }\textbf {\bibinfo {volume} {94}},\ \bibinfo {eid} {064038} (\bibinfo
  {year} {2016})},\ \Eprint {http://arxiv.org/abs/1512.05232} {arXiv:1512.05232
  [gr-qc]} \BibitemShut {NoStop}%
\bibitem [{\citenamefont {{Minazzoli}}()}]{minazzoli:2020ds}%
  \BibitemOpen
  \bibfield  {author} {\bibinfo {author} {\bibfnamefont {Olivier}\ \bibnamefont
  {{Minazzoli}}},\ }\bibfield  {title} {\enquote {\bibinfo {title} {{De Sitter
  space-times in Entangled Relativity}},}\ }\href@noop {} {\bibinfo  {journal}
  {in preparation}\ }\BibitemShut {NoStop}%
\bibitem [{\citenamefont {{Minazzoli}}\ and\ \citenamefont
  {{Santos}}()}]{minazzoli:2020bh}%
  \BibitemOpen
\bibfield  {journal} {  }\bibfield  {author} {\bibinfo {author} {\bibfnamefont
  {Olivier}\ \bibnamefont {{Minazzoli}}}\ and\ \bibinfo {author} {\bibfnamefont
  {Edison}\ \bibnamefont {{Santos}}},\ }\bibfield  {title} {\enquote {\bibinfo
  {title} {{External Schwarzschild metric as a good approximation of spherical
  black hole solutions in Entangled Relativity}},}\ }\href@noop {} {\bibinfo
  {journal} {in preparation}\ }\BibitemShut {NoStop}%
\bibitem [{\citenamefont {{Wald}}(1984)}]{wald:1984bk}%
  \BibitemOpen
\bibfield  {journal} {  }\bibfield  {author} {\bibinfo {author} {\bibfnamefont
  {R.~M.}\ \bibnamefont {{Wald}}},\ }\href@noop {} {\emph {\bibinfo {title}
  {{General relativity}}}}\ (\bibinfo {year} {1984})\BibitemShut {NoStop}%
\bibitem [{cod()}]{code}%
  \BibitemOpen
  \href@noop {} {}\bibinfo {howpublished} {Code and script to compute all the
  figures,
  \url{https://github.com/olivierrousselle/TOV-dilaton/tree/integration}}\BibitemShut
  {NoStop}%
\bibitem [{\citenamefont {{Schl{\"o}gel}}\ \emph {et~al.}(2014)\citenamefont
  {{Schl{\"o}gel}}, \citenamefont {{Rinaldi}}, \citenamefont {{Staelens}},\
  and\ \citenamefont {{F{\"u}zfa}}}]{schlogel:2014pr}%
  \BibitemOpen
  \bibfield  {author} {\bibinfo {author} {\bibfnamefont {Sandrine}\
  \bibnamefont {{Schl{\"o}gel}}}, \bibinfo {author} {\bibfnamefont
  {Massimiliano}\ \bibnamefont {{Rinaldi}}}, \bibinfo {author} {\bibfnamefont
  {Fran{\c{c}}ois}\ \bibnamefont {{Staelens}}}, \ and\ \bibinfo {author}
  {\bibfnamefont {Andr{\'e}}\ \bibnamefont {{F{\"u}zfa}}},\ }\bibfield  {title}
  {\enquote {\bibinfo {title} {{Particlelike solutions in modified gravity: The
  Higgs monopole}},}\ }\href {\doibase 10.1103/PhysRevD.90.044056} {\bibfield
  {journal} {\bibinfo  {journal} {\prd}\ }\textbf {\bibinfo {volume} {90}},\
  \bibinfo {eid} {044056} (\bibinfo {year} {2014})},\ \Eprint
  {http://arxiv.org/abs/1405.5476} {arXiv:1405.5476 [gr-qc]} \BibitemShut
  {NoStop}%
\bibitem [{\citenamefont {{Minazzoli}}\ and\ \citenamefont
  {{Harko}}(2012)}]{minazzoli:2012pr}%
  \BibitemOpen
  \bibfield  {author} {\bibinfo {author} {\bibfnamefont {Olivier}\ \bibnamefont
  {{Minazzoli}}}\ and\ \bibinfo {author} {\bibfnamefont {Tiberiu}\ \bibnamefont
  {{Harko}}},\ }\bibfield  {title} {\enquote {\bibinfo {title} {{New derivation
  of the Lagrangian of a perfect fluid with a barotropic equation of state}},}\
  }\href {\doibase 10.1103/PhysRevD.86.087502} {\bibfield  {journal} {\bibinfo
  {journal} {\prd}\ }\textbf {\bibinfo {volume} {86}},\ \bibinfo {eid} {087502}
  (\bibinfo {year} {2012})},\ \Eprint {http://arxiv.org/abs/1209.2754}
  {arXiv:1209.2754 [gr-qc]} \BibitemShut {NoStop}%
\bibitem [{\citenamefont {{Minazzoli}}(2013)}]{minazzoli:2013pd}%
  \BibitemOpen
  \bibfield  {author} {\bibinfo {author} {\bibfnamefont {Olivier}\ \bibnamefont
  {{Minazzoli}}},\ }\bibfield  {title} {\enquote {\bibinfo {title}
  {{Conservation laws in theories with universal gravity/matter coupling}},}\
  }\href {\doibase 10.1103/PhysRevD.88.027506} {\bibfield  {journal} {\bibinfo
  {journal} {\prd}\ }\textbf {\bibinfo {volume} {88}},\ \bibinfo {eid} {027506}
  (\bibinfo {year} {2013})},\ \Eprint {http://arxiv.org/abs/1307.1590}
  {arXiv:1307.1590 [gr-qc]} \BibitemShut {NoStop}%
\bibitem [{\citenamefont {Avelino}\ and\ \citenamefont
  {Azevedo}(2018)}]{avelino:2018pr}%
  \BibitemOpen
  \bibfield  {author} {\bibinfo {author} {\bibfnamefont {P.~P.}\ \bibnamefont
  {Avelino}}\ and\ \bibinfo {author} {\bibfnamefont {R.~P.~L.}\ \bibnamefont
  {Azevedo}},\ }\bibfield  {title} {\enquote {\bibinfo {title} {Perfect fluid
  lagrangian and its cosmological implications in theories of gravity with
  nonminimally coupled matter fields},}\ }\href {\doibase
  10.1103/PhysRevD.97.064018} {\bibfield  {journal} {\bibinfo  {journal} {Phys.
  Rev. D}\ }\textbf {\bibinfo {volume} {97}},\ \bibinfo {pages} {064018}
  (\bibinfo {year} {2018})}\BibitemShut {NoStop}%
\bibitem [{\citenamefont {Avelino}\ and\ \citenamefont
  {Sousa}(2018)}]{avelino:2018pd}%
  \BibitemOpen
  \bibfield  {author} {\bibinfo {author} {\bibfnamefont {P.~P.}\ \bibnamefont
  {Avelino}}\ and\ \bibinfo {author} {\bibfnamefont {L.}~\bibnamefont
  {Sousa}},\ }\bibfield  {title} {\enquote {\bibinfo {title} {Matter lagrangian
  of particles and fluids},}\ }\href {\doibase 10.1103/PhysRevD.97.064019}
  {\bibfield  {journal} {\bibinfo  {journal} {Phys. Rev. D}\ }\textbf {\bibinfo
  {volume} {97}},\ \bibinfo {pages} {064019} (\bibinfo {year}
  {2018})}\BibitemShut {NoStop}%
\bibitem [{\citenamefont {{Thorsrud}}\ \emph {et~al.}(2012)\citenamefont
  {{Thorsrud}}, \citenamefont {{Mota}},\ and\ \citenamefont
  {{Hervik}}}]{thorsrud:2012jh}%
  \BibitemOpen
  \bibfield  {author} {\bibinfo {author} {\bibfnamefont {Mikjel}\ \bibnamefont
  {{Thorsrud}}}, \bibinfo {author} {\bibfnamefont {David~F.}\ \bibnamefont
  {{Mota}}}, \ and\ \bibinfo {author} {\bibfnamefont {Sigbj{\o}rn}\
  \bibnamefont {{Hervik}}},\ }\bibfield  {title} {\enquote {\bibinfo {title}
  {{Cosmology of a scalar field coupled to matter and an isotropy-violating
  Maxwell field}},}\ }\href {\doibase 10.1007/JHEP10(2012)066} {\bibfield
  {journal} {\bibinfo  {journal} {Journal of High Energy Physics}\ }\textbf
  {\bibinfo {volume} {2012}},\ \bibinfo {eid} {66} (\bibinfo {year} {2012})},\
  \Eprint {http://arxiv.org/abs/1205.6261} {arXiv:1205.6261 [hep-th]}
  \BibitemShut {NoStop}%
\bibitem [{\citenamefont {Turner}(1983)}]{turner:1983pr}%
  \BibitemOpen
  \bibfield  {author} {\bibinfo {author} {\bibfnamefont {Michael~S.}\
  \bibnamefont {Turner}},\ }\bibfield  {title} {\enquote {\bibinfo {title}
  {Coherent scalar-field oscillations in an expanding universe},}\ }\href
  {\doibase 10.1103/PhysRevD.28.1243} {\bibfield  {journal} {\bibinfo
  {journal} {Phys. Rev. D}\ }\textbf {\bibinfo {volume} {28}},\ \bibinfo
  {pages} {1243--1247} (\bibinfo {year} {1983})}\BibitemShut {NoStop}%
\bibitem [{\citenamefont {{Alsing}}\ \emph {et~al.}(2018)\citenamefont
  {{Alsing}}, \citenamefont {{Silva}},\ and\ \citenamefont
  {{Berti}}}]{alsing:2018mn}%
  \BibitemOpen
  \bibfield  {author} {\bibinfo {author} {\bibfnamefont {Justin}\ \bibnamefont
  {{Alsing}}}, \bibinfo {author} {\bibfnamefont {Hector~O.}\ \bibnamefont
  {{Silva}}}, \ and\ \bibinfo {author} {\bibfnamefont {Emanuele}\ \bibnamefont
  {{Berti}}},\ }\bibfield  {title} {\enquote {\bibinfo {title} {{Evidence for a
  maximum mass cut-off in the neutron star mass distribution and constraints on
  the equation of state}},}\ }\href {\doibase 10.1093/mnras/sty1065} {\bibfield
   {journal} {\bibinfo  {journal} {\mnras}\ }\textbf {\bibinfo {volume}
  {478}},\ \bibinfo {pages} {1377--1391} (\bibinfo {year} {2018})},\ \Eprint
  {http://arxiv.org/abs/1709.07889} {arXiv:1709.07889 [astro-ph.HE]}
  \BibitemShut {NoStop}%
\bibitem [{\citenamefont {{Biswas}}\ \emph {et~al.}(2020)\citenamefont
  {{Biswas}}, \citenamefont {{Char}}, \citenamefont {{Nandi}},\ and\
  \citenamefont {{Bose}}}]{biswas:2020ar}%
  \BibitemOpen
  \bibfield  {author} {\bibinfo {author} {\bibfnamefont {Bhaskar}\ \bibnamefont
  {{Biswas}}}, \bibinfo {author} {\bibfnamefont {Prasanta}\ \bibnamefont
  {{Char}}}, \bibinfo {author} {\bibfnamefont {Rana}\ \bibnamefont {{Nandi}}},
  \ and\ \bibinfo {author} {\bibfnamefont {Sukanta}\ \bibnamefont {{Bose}}},\
  }\bibfield  {title} {\enquote {\bibinfo {title} {{Hint of a tension between
  Nuclear physics and Astrophysical observations}},}\ }\href@noop {} {\bibfield
   {journal} {\bibinfo  {journal} {arXiv e-prints}\ ,\ \bibinfo {eid}
  {arXiv:2008.01582}} (\bibinfo {year} {2020})},\ \Eprint
  {http://arxiv.org/abs/2008.01582} {arXiv:2008.01582 [astro-ph.HE]}
  \BibitemShut {NoStop}%
\bibitem [{\citenamefont {{Demorest}}\ \emph {et~al.}(2010)\citenamefont
  {{Demorest}}, \citenamefont {{Pennucci}}, \citenamefont {{Ransom}},
  \citenamefont {{Roberts}},\ and\ \citenamefont
  {{Hessels}}}]{demorest:2010na}%
  \BibitemOpen
  \bibfield  {author} {\bibinfo {author} {\bibfnamefont {P.~B.}\ \bibnamefont
  {{Demorest}}}, \bibinfo {author} {\bibfnamefont {T.}~\bibnamefont
  {{Pennucci}}}, \bibinfo {author} {\bibfnamefont {S.~M.}\ \bibnamefont
  {{Ransom}}}, \bibinfo {author} {\bibfnamefont {M.~S.~E.}\ \bibnamefont
  {{Roberts}}}, \ and\ \bibinfo {author} {\bibfnamefont {J.~W.~T.}\
  \bibnamefont {{Hessels}}},\ }\bibfield  {title} {\enquote {\bibinfo {title}
  {{A two-solar-mass neutron star measured using Shapiro delay}},}\ }\href
  {\doibase 10.1038/nature09466} {\bibfield  {journal} {\bibinfo  {journal}
  {\nat}\ }\textbf {\bibinfo {volume} {467}},\ \bibinfo {pages} {1081--1083}
  (\bibinfo {year} {2010})},\ \Eprint {http://arxiv.org/abs/1010.5788}
  {arXiv:1010.5788 [astro-ph.HE]} \BibitemShut {NoStop}%
\bibitem [{\citenamefont {{Fonseca}}\ \emph {et~al.}(2016)\citenamefont
  {{Fonseca}}, \citenamefont {{Pennucci}}, \citenamefont {{Ellis}},
  \citenamefont {{Stairs}}, \citenamefont {{Nice}}, \citenamefont {{Ransom}},
  \citenamefont {{Demorest}}, \citenamefont {{Arzoumanian}}, \citenamefont
  {{Crowter}}, \citenamefont {{Dolch}}, \citenamefont {{Ferdman}},
  \citenamefont {{Gonzalez}}, \citenamefont {{Jones}}, \citenamefont {{Jones}},
  \citenamefont {{Lam}}, \citenamefont {{Levin}}, \citenamefont {{McLaughlin}},
  \citenamefont {{Stovall}}, \citenamefont {{Swiggum}},\ and\ \citenamefont
  {{Zhu}}}]{fonseca:2016aj}%
  \BibitemOpen
  \bibfield  {author} {\bibinfo {author} {\bibfnamefont {Emmanuel}\
  \bibnamefont {{Fonseca}}}, \bibinfo {author} {\bibfnamefont {Timothy~T.}\
  \bibnamefont {{Pennucci}}}, \bibinfo {author} {\bibfnamefont {Justin~A.}\
  \bibnamefont {{Ellis}}}, \bibinfo {author} {\bibfnamefont {Ingrid~H.}\
  \bibnamefont {{Stairs}}}, \bibinfo {author} {\bibfnamefont {David~J.}\
  \bibnamefont {{Nice}}}, \bibinfo {author} {\bibfnamefont {Scott~M.}\
  \bibnamefont {{Ransom}}}, \bibinfo {author} {\bibfnamefont {Paul~B.}\
  \bibnamefont {{Demorest}}}, \bibinfo {author} {\bibfnamefont {Zaven}\
  \bibnamefont {{Arzoumanian}}}, \bibinfo {author} {\bibfnamefont {Kathryn}\
  \bibnamefont {{Crowter}}}, \bibinfo {author} {\bibfnamefont {Timothy}\
  \bibnamefont {{Dolch}}}, \bibinfo {author} {\bibfnamefont {Robert~D.}\
  \bibnamefont {{Ferdman}}}, \bibinfo {author} {\bibfnamefont {Marjorie~E.}\
  \bibnamefont {{Gonzalez}}}, \bibinfo {author} {\bibfnamefont {Glenn}\
  \bibnamefont {{Jones}}}, \bibinfo {author} {\bibfnamefont {Megan~L.}\
  \bibnamefont {{Jones}}}, \bibinfo {author} {\bibfnamefont {Michael~T.}\
  \bibnamefont {{Lam}}}, \bibinfo {author} {\bibfnamefont {Lina}\ \bibnamefont
  {{Levin}}}, \bibinfo {author} {\bibfnamefont {Maura~A.}\ \bibnamefont
  {{McLaughlin}}}, \bibinfo {author} {\bibfnamefont {Kevin}\ \bibnamefont
  {{Stovall}}}, \bibinfo {author} {\bibfnamefont {Joseph~K.}\ \bibnamefont
  {{Swiggum}}}, \ and\ \bibinfo {author} {\bibfnamefont {Weiwei}\ \bibnamefont
  {{Zhu}}},\ }\bibfield  {title} {\enquote {\bibinfo {title} {{The NANOGrav
  Nine-year Data Set: Mass and Geometric Measurements of Binary Millisecond
  Pulsars}},}\ }\href {\doibase 10.3847/0004-637X/832/2/167} {\bibfield
  {journal} {\bibinfo  {journal} {\apj}\ }\textbf {\bibinfo {volume} {832}},\
  \bibinfo {eid} {167} (\bibinfo {year} {2016})},\ \Eprint
  {http://arxiv.org/abs/1603.00545} {arXiv:1603.00545 [astro-ph.HE]}
  \BibitemShut {NoStop}%
\bibitem [{\citenamefont {{Cromartie}}\ \emph {et~al.}(2020)\citenamefont
  {{Cromartie}}, \citenamefont {{Fonseca}}, \citenamefont {{Ransom}},
  \citenamefont {{Demorest}}, \citenamefont {{Arzoumanian}}, \citenamefont
  {{Blumer}}, \citenamefont {{Brook}}, \citenamefont {{DeCesar}}, \citenamefont
  {{Dolch}}, \citenamefont {{Ellis}}, \citenamefont {{Ferdman}}, \citenamefont
  {{Ferrara}}, \citenamefont {{Garver-Daniels}}, \citenamefont {{Gentile}},
  \citenamefont {{Jones}}, \citenamefont {{Lam}}, \citenamefont {{Lorimer}},
  \citenamefont {{Lynch}}, \citenamefont {{McLaughlin}}, \citenamefont {{Ng}},
  \citenamefont {{Nice}}, \citenamefont {{Pennucci}}, \citenamefont
  {{Spiewak}}, \citenamefont {{Stairs}}, \citenamefont {{Stovall}},
  \citenamefont {{Swiggum}},\ and\ \citenamefont {{Zhu}}}]{cromartie:2020na}%
  \BibitemOpen
  \bibfield  {author} {\bibinfo {author} {\bibfnamefont {H.~T.}\ \bibnamefont
  {{Cromartie}}}, \bibinfo {author} {\bibfnamefont {E.}~\bibnamefont
  {{Fonseca}}}, \bibinfo {author} {\bibfnamefont {S.~M.}\ \bibnamefont
  {{Ransom}}}, \bibinfo {author} {\bibfnamefont {P.~B.}\ \bibnamefont
  {{Demorest}}}, \bibinfo {author} {\bibfnamefont {Z.}~\bibnamefont
  {{Arzoumanian}}}, \bibinfo {author} {\bibfnamefont {H.}~\bibnamefont
  {{Blumer}}}, \bibinfo {author} {\bibfnamefont {P.~R.}\ \bibnamefont
  {{Brook}}}, \bibinfo {author} {\bibfnamefont {M.~E.}\ \bibnamefont
  {{DeCesar}}}, \bibinfo {author} {\bibfnamefont {T.}~\bibnamefont {{Dolch}}},
  \bibinfo {author} {\bibfnamefont {J.~A.}\ \bibnamefont {{Ellis}}}, \bibinfo
  {author} {\bibfnamefont {R.~D.}\ \bibnamefont {{Ferdman}}}, \bibinfo {author}
  {\bibfnamefont {E.~C.}\ \bibnamefont {{Ferrara}}}, \bibinfo {author}
  {\bibfnamefont {N.}~\bibnamefont {{Garver-Daniels}}}, \bibinfo {author}
  {\bibfnamefont {P.~A.}\ \bibnamefont {{Gentile}}}, \bibinfo {author}
  {\bibfnamefont {M.~L.}\ \bibnamefont {{Jones}}}, \bibinfo {author}
  {\bibfnamefont {M.~T.}\ \bibnamefont {{Lam}}}, \bibinfo {author}
  {\bibfnamefont {D.~R.}\ \bibnamefont {{Lorimer}}}, \bibinfo {author}
  {\bibfnamefont {R.~S.}\ \bibnamefont {{Lynch}}}, \bibinfo {author}
  {\bibfnamefont {M.~A.}\ \bibnamefont {{McLaughlin}}}, \bibinfo {author}
  {\bibfnamefont {C.}~\bibnamefont {{Ng}}}, \bibinfo {author} {\bibfnamefont
  {D.~J.}\ \bibnamefont {{Nice}}}, \bibinfo {author} {\bibfnamefont {T.~T.}\
  \bibnamefont {{Pennucci}}}, \bibinfo {author} {\bibfnamefont
  {R.}~\bibnamefont {{Spiewak}}}, \bibinfo {author} {\bibfnamefont {I.~H.}\
  \bibnamefont {{Stairs}}}, \bibinfo {author} {\bibfnamefont {K.}~\bibnamefont
  {{Stovall}}}, \bibinfo {author} {\bibfnamefont {J.~K.}\ \bibnamefont
  {{Swiggum}}}, \ and\ \bibinfo {author} {\bibfnamefont {W.~W.}\ \bibnamefont
  {{Zhu}}},\ }\bibfield  {title} {\enquote {\bibinfo {title} {{Relativistic
  Shapiro delay measurements of an extremely massive millisecond pulsar}},}\
  }\href {\doibase 10.1038/s41550-019-0880-2} {\bibfield  {journal} {\bibinfo
  {journal} {Nature Astronomy}\ }\textbf {\bibinfo {volume} {4}},\ \bibinfo
  {pages} {72--76} (\bibinfo {year} {2020})},\ \Eprint
  {http://arxiv.org/abs/1904.06759} {arXiv:1904.06759 [astro-ph.HE]}
  \BibitemShut {NoStop}%
\bibitem [{\citenamefont {{Gendreau}}\ and\ \citenamefont
  {{Arzoumanian}}(2017)}]{gendreau:2017na}%
  \BibitemOpen
  \bibfield  {author} {\bibinfo {author} {\bibfnamefont {Keith}\ \bibnamefont
  {{Gendreau}}}\ and\ \bibinfo {author} {\bibfnamefont {Zaven}\ \bibnamefont
  {{Arzoumanian}}},\ }\bibfield  {title} {\enquote {\bibinfo {title}
  {{Searching for a pulse}},}\ }\href {\doibase 10.1038/s41550-017-0301-3}
  {\bibfield  {journal} {\bibinfo  {journal} {Nature Astronomy}\ }\textbf
  {\bibinfo {volume} {1}},\ \bibinfo {pages} {895--895} (\bibinfo {year}
  {2017})}\BibitemShut {NoStop}%
\bibitem [{\citenamefont {{Silva}}\ and\ \citenamefont
  {{Yunes}}(2019{\natexlab{a}})}]{silva:2019pr}%
  \BibitemOpen
  \bibfield  {author} {\bibinfo {author} {\bibfnamefont {Hector~O.}\
  \bibnamefont {{Silva}}}\ and\ \bibinfo {author} {\bibfnamefont {Nicol{\'a}s}\
  \bibnamefont {{Yunes}}},\ }\bibfield  {title} {\enquote {\bibinfo {title}
  {{Neutron star pulse profiles in scalar-tensor theories of gravity}},}\
  }\href {\doibase 10.1103/PhysRevD.99.044034} {\bibfield  {journal} {\bibinfo
  {journal} {\prd}\ }\textbf {\bibinfo {volume} {99}},\ \bibinfo {eid} {044034}
  (\bibinfo {year} {2019}{\natexlab{a}})},\ \Eprint
  {http://arxiv.org/abs/1808.04391} {arXiv:1808.04391 [gr-qc]} \BibitemShut
  {NoStop}%
\bibitem [{\citenamefont {{Silva}}\ and\ \citenamefont
  {{Yunes}}(2019{\natexlab{b}})}]{silva:2019cq}%
  \BibitemOpen
  \bibfield  {author} {\bibinfo {author} {\bibfnamefont {Hector~O.}\
  \bibnamefont {{Silva}}}\ and\ \bibinfo {author} {\bibfnamefont {Nicol{\'a}s}\
  \bibnamefont {{Yunes}}},\ }\bibfield  {title} {\enquote {\bibinfo {title}
  {{Neutron star pulse profile observations as extreme gravity probes}},}\
  }\href {\doibase 10.1088/1361-6382/ab3560} {\bibfield  {journal} {\bibinfo
  {journal} {Classical and Quantum Gravity}\ }\textbf {\bibinfo {volume}
  {36}},\ \bibinfo {eid} {17LT01} (\bibinfo {year} {2019}{\natexlab{b}})},\
  \Eprint {http://arxiv.org/abs/1902.10269} {arXiv:1902.10269 [gr-qc]}
  \BibitemShut {NoStop}%
\bibitem [{\citenamefont {{Xu}}\ \emph {et~al.}(2020)\citenamefont {{Xu}},
  \citenamefont {{Gao}},\ and\ \citenamefont {{Shao}}}]{xu:2020pr}%
  \BibitemOpen
  \bibfield  {author} {\bibinfo {author} {\bibfnamefont {Rui}\ \bibnamefont
  {{Xu}}}, \bibinfo {author} {\bibfnamefont {Yong}\ \bibnamefont {{Gao}}}, \
  and\ \bibinfo {author} {\bibfnamefont {Lijing}\ \bibnamefont {{Shao}}},\
  }\bibfield  {title} {\enquote {\bibinfo {title} {{Strong-field effects in
  massive scalar-tensor gravity for slowly spinning neutron stars and
  application to x-ray pulsar pulse profiles}},}\ }\href {\doibase
  10.1103/PhysRevD.102.064057} {\bibfield  {journal} {\bibinfo  {journal}
  {\prd}\ }\textbf {\bibinfo {volume} {102}},\ \bibinfo {eid} {064057}
  (\bibinfo {year} {2020})},\ \Eprint {http://arxiv.org/abs/2007.10080}
  {arXiv:2007.10080 [gr-qc]} \BibitemShut {NoStop}%
\bibitem [{\citenamefont {{Abbott et al.}}\ \emph {et~al.}(2017)\citenamefont
  {{Abbott et al.}}, \citenamefont {{LIGO Scientific Collaboration}},\ and\
  \citenamefont {{Virgo Collaboration}}}]{abbott:2017pl}%
  \BibitemOpen
  \bibfield  {author} {\bibinfo {author} {\bibfnamefont {B.~P.}\ \bibnamefont
  {{Abbott et al.}}}, \bibinfo {author} {\bibnamefont {{LIGO Scientific
  Collaboration}}}, \ and\ \bibinfo {author} {\bibnamefont {{Virgo
  Collaboration}}},\ }\bibfield  {title} {\enquote {\bibinfo {title}
  {{GW170817: Observation of Gravitational Waves from a Binary Neutron Star
  Inspiral}},}\ }\href {\doibase 10.1103/PhysRevLett.119.161101} {\bibfield
  {journal} {\bibinfo  {journal} {\prl}\ }\textbf {\bibinfo {volume} {119}},\
  \bibinfo {eid} {161101} (\bibinfo {year} {2017})},\ \Eprint
  {http://arxiv.org/abs/1710.05832} {arXiv:1710.05832 [gr-qc]} \BibitemShut
  {NoStop}%
\end{thebibliography}%

\appendix

\section{Perfect fluid Lagrangian and conservation of the rest-mass density}
\label{app:cons}

In this section, we slightly modify a demonstration that can already be found in \cite{minazzoli:2013pd}. Let us assume that the rest-mass density is not conserved, such that
\be
\label{eq:nonconsrho0}
\nabla_\sigma(\rho_0 u^\sigma) = D,
\ee
where $D$ is a scalar to be determined. Furthermore, one uses the usual definition of the total energy density, given in terms of the rest-mass enegy density and pressure
\be
\rho \equiv \rho_0\left(c^{2}-\frac{P}{\rho_0}+\int \frac{\mathrm{d} P}{\rho_0}\right).
\ee
Injecting this definition into Eq. (\ref{eq:nonconsrho0}), one gets
\be
\nabla_{\sigma}\left(\rho u^{\sigma}\right)=\left(\frac{\rho+P}{\rho_0}\right) D-P \nabla_{\sigma} u^{\sigma}.
\ee
Now, let us write the effective on-shell matter Lagrangian as 
\be
\label{eq:defLgen}
\mathcal{L}_m = - \alpha \rho + \beta P,
\ee
where $\alpha$ and $\beta$ are merely constants that are used in order to parametrize the three potential cases considered $\mathcal{L}_m = - \rho$, $P$ or $T$--- respectively $(\alpha = 1, \beta=0)$,$(\alpha = 0, \beta=1)$ or $(\alpha = 1, \beta=3)$. Taking the divergence of the perfect fluid stress-energy tensor (\ref{eq:tab}), one gets
\be
\begin{aligned}
\nabla_{\sigma} T^{\mu \sigma}=&(\rho+P) u^{\sigma} \nabla_{\sigma} u^{\mu}+\left(g^{\mu \sigma}+u^{\mu} u^{\sigma}\right) \nabla_{\sigma} P \\
&+u^{\mu}\left(\frac{\rho+P}{\rho_0}\right) D.
\end{aligned}
\ee
On the other side, the conservation equation (\ref{eq:noncons}), together with Eq. (\ref{eq:defLgen}) and the perfect fluid stress-energy tensor (\ref{eq:tab}) leads to
\be
\begin{aligned}
\nabla_{\sigma} T^{\mu \sigma}=&- (\rho+P)\left[g^{\mu \sigma}+u^{\mu} u^{\sigma}\right] \partial_{\sigma} \ln \sqrt{\phi} \\
&+\left[(1-\alpha) \rho + \beta P\right] g^{\mu \sigma} \partial_{\sigma} \ln \sqrt{\phi}.
\end{aligned}
\ee
Multiplying by $u_\mu$ the previous two equations and equating them gives
\be
D = -\frac{\rho_0}{\rho+P}\left[(1-\alpha) \rho + \beta P\right] u^{\sigma} \partial_{\sigma} \ln\sqrt{\phi}.
\ee
Hence, one has $D(\alpha=1, \beta =0) = 0$ in Eq. (\ref{eq:nonconsrho0}), which corresponds to $\mathcal{L}_m = -\rho$ in Eq. (\ref{eq:defLgen}). Therefore, the conservation of the rest-mass energy density corresponds to a fluid that is such that its on-shell Lagrangian is $\mathcal{L}_m = -\rho$---which is the main assumption in this work.

\end{document}